\documentclass[twocolumn,superscriptaddress,prd]{revtex4}
\usepackage{graphicx}
\usepackage{dcolumn}
\usepackage{bm}
\usepackage{amsmath}
\usepackage{amsfonts}

\begin{document}

\title{On Born-Infeld Gravity in Weitzenb\"{o}ck spacetime}
\author{Rafael Ferraro}
\email{ferraro@iafe.uba.ar}
\thanks{Member of Carrera del Investigador Cient\'{\i}fico (CONICET,
Argentina)}
\affiliation{Instituto de Astronom\'\i a y F\'\i sica del Espacio, Casilla de Correo 67,
Sucursal 28, 1428 Buenos Aires, Argentina}
\affiliation{Departamento de F\'\i sica, Facultad de Ciencias Exactas y Naturales,
Universidad de Buenos Aires, Ciudad Universitaria, Pabell\'on I, 1428 Buenos
Aires, Argentina}
\author{Franco Fiorini}
\email{franco@iafe.uba.ar}
\affiliation{Instituto de Astronom\'\i a y F\'\i sica del Espacio, Casilla de Correo 67,
Sucursal 28, 1428 Buenos Aires, Argentina}

\begin{abstract}
Using the Teleparallel Equivalent of General Relativity formulated
in Weitzenb\"{o}ck spacetime, we thoroughly explore a kind of
Born-Infeld regular gravity leading to second order field
equations for the vielbein components. We explicitly solve the
equations of motion for two examples: the extended BTZ black hole,
which results to exist even if the cosmological constant is
positive, and a cosmological model with matter, where the scale
factor results to be well behaved, giving so a singularity-free
solution.
\end{abstract}

 \maketitle



\section{Introduction: ultraviolet corrections to GR}

In the last decade a wide variety of modified theories of gravity has been
studied with the aim of solving or smoothing some puzzling features of
conventional gravity and cosmology such as singularities, particle horizons,
acceleration of the universe expansion, etc. Many of these modified theories
of gravity consist in the mere deformation of the current theory. In this
case, one starts from a known Lagrangian $\mathcal{L}=e\,L$, where $L$ is
invariant and $e$ is a density under general coordinate changes, and then
the theory is deformed by replacing the Lagrangian by $\mathcal{L}%
_{D}=e\,f(L).$ It is expected that a suitable choice of the
function $f$ will heal the unwanted features of the original
theory. To explain the method, let us consider an invariant
Lagrangian $L=L(\phi ^{a},\phi _{,\mu }^{a}\ ,\phi _{,\mu \nu \
}^{a},...,\,x^{\mu })$ and a density $e$ that does not depend on
the derivatives of the fields $\phi ^{a}$: $e=e(\phi ^{a},x^{\mu
})$ (this is because $\phi ^{a}$ will later become a field
describing the geometry, and so the density $e$ will be the square
root of the determinant of the metric). Thus the Euler-Lagrange
equations for the deformed Lagrangian $\mathcal{L}_{D}=e\,f(L)$
are
\begin{eqnarray}
0 &=&...-\partial _{\mu }\,\partial _{\nu }\left( \frac{\partial \mathcal{L}%
_{D}}{\partial \phi _{,\mu \nu \ }^{a}}\right) +\partial _{\mu }\left( \frac{%
\partial \mathcal{L}_{D}}{\partial \phi _{,\mu }^{a}}\right) -\frac{\partial
\mathcal{L}_{D}}{\partial \phi ^{a}}=  \notag \\
&&...-\partial _{\mu }\,\partial _{\nu }\left( f^{\ \prime }(L)\ \frac{%
\partial \mathcal{L}}{\partial \phi _{,\mu \nu \ }^{a}}\right) +\partial
_{\mu }\left( f^{\ \prime }(L)\ \frac{\partial \mathcal{L}}{\partial \phi
_{,\mu }^{a}}\right)   \notag \\
&&-f^{\ \prime }(L)\ \frac{\partial \mathcal{L}}{\partial \phi
^{a}}+\left( L\ f^{\ \prime }(L)-f(L)\right) \ \frac{\partial
e}{\partial \phi ^{a}}. \label{lagrangeeq}
\end{eqnarray}%
If the deformed Lagrangian is intended to modify only the strong
field (large $L$) regime, then $f$ should satisfy
\begin{equation}
f(L)\simeq L+O(L^{2}),  \label{ldebil}
\end{equation}%
i.e.,
\begin{equation}
f(0)=0,\hspace{2cm}f^{\ \prime }(0)=1.  \label{fdebil}
\end{equation}%
In general, equations (\ref{lagrangeeq}) will have solutions
differing from those coming from the original Lagrangian
$\mathcal{L}=e\,L.$ However, it should be noted that not all the
solutions of the original theory get deformed by this procedure.
In fact, let us consider solutions of the original theory such
that $L=0$. In this case, by substituting $L=0$ in
(\ref{lagrangeeq}) and using (\ref{fdebil}) it results that the
last term vanishes. Moreover, since $f^{\ \prime }(0)=1$, then
those solutions of the original theory having $L=0$ also solve the
Euler-Lagrange equations for the deformed Lagrangian
$\mathcal{L}_{D}$. In particular, the invariant $L$ for general
relativity (\textbf{GR}) is the curvature scalar $R$ associated
with the Levi-Civita connection, which is null for all the
(vacuum) solutions. Thus general relativity is a quite rigid
theory, because its vacuum solutions remain as solutions for the
(vacuum) field equations of deformed theories
$\mathcal{L}_{D}\propto\sqrt{-g}\,f(R)$, with $f$ satisfying
conditions (\ref{fdebil}). This is a rather singular feature which
is not shared by other field theories. For instance, in Maxwell
electromagnetism it is $L\propto E^{2}-B^{2}$, and only some
vacuum solutions -mainly plane waves- make null the Maxwell
Lagrangian. \bigskip

Contrasting with other theories, general relativity Lagrangian
$L\propto R$ contains second derivatives of the metric. In spite
of this feature, Einstein equations result to be second order
because the fourth order terms in Euler-Lagrange equations cancel
out (in other words, second derivatives in $L$ appear just
contributing to a divergence term in the action). This property is
lost in the deformed theory
$\mathcal{L}_{D}\propto\sqrt{-g}\,f(R)$, whose dynamical equations
become fourth order, as it follows from Eq.$\;$(\ref{lagrangeeq}).
This undesirable fact is usually relieved by splitting the metric
in a new metric tensor times a conformal factor depending on a
scalar field; the scalar field becomes a constant when the ($f^{\
\prime }=1$) general relativity case is retrieved. This procedure
allows to reformulate a $f(R)$ theory as a Brans-Dicke-like
scalar-tensor theory of gravity having $\omega =0$ (metric
formalism \cite{teys,chiba}) or $\omega =-3/2$ (Palatini formalism
\cite{hamity,vollick,olmo}); thus the new metric results to be
governed by second order equations and the extra degrees of
freedom are placed in a scalar field fulfilling a second order
equation too. However, the scalar-tensor reformulation of $f(R)$
theories results in violations of the weak equivalence principle,
since matter and gravity would couple not only through the (new)
metric but also through the scalar field \cite{brans,olmo2}.
Incidentally, we mention that not all the $f(R)$'s appearing in
the literature fulfill the conditions (\ref{fdebil}); see, for
instance, the $f(R)$ used to build the spherically symmetric
vacuum solution in Ref.$\;$\cite{clifton}, or the one proposed in
Ref.$\;$\cite{carroll,vollick} to explain the acceleration of the
universe as an effect of modified gravity at the low curvature
regime (which, if regarded as a Brans-Dicke-like theory, can be
dismissed on the basis of well established post-Newtonian
constraints \cite{chiba,olmo,olmo3}; the Newtonian limit in
Palatini formalism is not retrieved either \cite{barraco}).

\bigskip

The problems inherent in the formulation of a $f(R)$ theory can be
avoided by starting from an alternative theory of gravity whose
Lagrangian only contains first derivatives of the dynamical
variables.  In a recent article \cite{Nos} we have proposed to
deform the teleparallel equivalent of general relativity
\textbf{TEGR} \cite{albert}. As currently formulated,
\textbf{TEGR} is an alternative formulation of general relativity.
Although the dynamical object of the theory is not the metric but
the vielbein $e_{\mu }^{a}(x)$, the
teleparallel action is invariant under local Lorentz transformations $%
\Lambda _{a}^{a^{\prime }}(x)$ of the vielbein,
\begin{equation}
e_{\mu }^{a}(x)\rightarrow e_{\mu }^{a^{\prime }}(x)=\Lambda
_{a}^{a^{\prime }}(x)\ e_{\mu }^{a}(x),
\end{equation}%
which do not change the metric
\begin{equation}
g_{\mu \nu }(x)=\eta _{ab}\ e_{\mu }^{a}(x)\ e_{\nu }^{b}(x),
\label{metric}
\end{equation}%
where $\eta _{ab}=diag(1,-1,-1,..)$. Since \textbf{TEGR} action is
not sensitive to some of the degrees of freedom of the vielbein,
the theory can be driven to be equivalent to general relativity
for the metric (\ref{metric}) \cite{Hayashi1,Pereira}.
Teleparallel Lagrangian is built from the torsion associated with
the Weitzenb\"{o}ck connection \cite{Weitz}
\begin{equation}
\overset{_{W}}{{\Gamma }}\!{_{\mu \nu }^{\lambda }}=e_{a}^{\lambda
}\,\partial _{\nu }e_{\mu }^{a}=-e_{\mu }^{a}\,\partial _{\nu
}e_{a}^{\lambda },  \label{Wei}
\end{equation}%
where $e_{a}^{\lambda }$ stands for the vielbein inverse matrix:
\begin{equation}
e_{a}^{\mu }\ e_{\mu }^{b}=\delta _{a}^{b}\ ,\ \ \ \ e_{a}^{\mu }\
e_{\nu }^{a}=\delta _{\nu }^{\mu}.
\end{equation}%
Weitzenb\"{o}ck connection has zero Riemann curvature
$\overset{_{W}}{R}$, but non-null torsion:
\begin{equation}
T_{\ \ \mu \nu }^{\lambda }=\overset{_{W}}{{\Gamma }}{_{\nu \mu }^{\lambda }}%
-\overset{_{W}}{{\Gamma }}{_{\mu \nu }^{\lambda }}=e_{a}^{\lambda
}(\partial _{\mu }e_{\nu }^{a}-\partial _{\nu }e_{\mu }^{a}).
\end{equation}%
The structure of the torsion tensor resembles the one of the
electromagnetic field tensor. Moreover, like Maxwell, teleparallel
Lagrangian density is quadratic in this tensor. In fact,
\textbf{TEGR} Lagrangian with cosmological constant $\Lambda $ is
\cite{Maluf}
\begin{equation}
\mathcal{L}_{\mathbf{T}}[e_{\mu }^{a}(x)]=\frac{e}{16\pi G}\
(\mathbb{S}\cdot \mathbb{T}-2\Lambda )  \label{lagrangianotel},
\end{equation}%
where $e$ is the determinant of matrix $e_{\mu }^{a}$ (which is equal to $%
\sqrt{-g}$), $\mathbb{S}\cdot \mathbb{T}\doteq S_{\rho }^{\ \ \mu \nu
}\,T_{\ \ \mu \nu }^{\rho }$, and $S_{\rho }^{\ \ \mu \nu }$ is defined as
\begin{eqnarray}
S_{\rho }^{\ \ \mu \nu } &=&-\frac{1}{4}\,(T_{\ \ \ \rho }^{\mu \nu }-T_{\ \
\ \rho }^{\nu \mu }-T_{\rho }^{\ \ \mu \nu })  \notag \\
&&+\frac{1}{2}\delta _{\rho }^{\mu }\,T_{\ \ \ \theta }^{\theta \nu }-\frac{1%
}{2}\delta _{\rho }^{\nu }\,T_{\ \ \ \theta }^{\theta \mu }.
\label{tensorS}
\end{eqnarray}%
While Einstein-Hilbert Lagrangian depends on second derivatives of the
metric, teleparallel Lagrangian is built just with first derivatives of the
vielbein, which makes more attractive the study of its deformation, in the
sense that the field equations of the deformed theory will remain being
second order equations. The Euler-Lagrange equations for the Lagrangian $%
\mathcal{L}_{\mathbf{T}}+\mathcal{L}_{\mathit{matter}}$ are
\begin{eqnarray}
\partial _{\sigma }(e\,e_{a}^{\lambda }\ S_{\lambda }^{\ \ \nu \sigma
})-e\;e_{a}^{\lambda }\ S_{\eta }^{\ \ \mu \nu }\,T_{\ \ \mu
\lambda
}^{\eta } &+&\frac{1}{4}\,e\;e_{a}^{\nu }\ \left( \mathbb{S}\cdot \mathbb{T}%
-2\Lambda \right)   \notag \\ &=&4\pi G\ e\,e_{a}^{\lambda }\
T_{\lambda }^{\ \nu },  \label{teleeq1}
\end{eqnarray}%
where $T_{\lambda }^{\ \nu }$ is the matter energy-momentum
tensor. By contracting Eq.$\;$(\ref{teleeq1}) with the inverse
vielbein $e_{\nu }^{a}$ one obtains for vacuum solutions
\begin{equation}
4\,e^{-1}\,e_{\nu }^{a}\ \partial _{\sigma }(e\,e_{a}^{\lambda }\
S_{\lambda }^{\ \ \nu \sigma })+(n-4)\,\mathbb{S}\cdot \mathbb{T}%
=2\,n\,\Lambda,   \label{teleeq2}
\end{equation}%
being $n$ the spacetime dimension. In contrast to general
relativity, where Einstein equations compels $R$ to vanish in
vacuum (or to be a constant when the cosmological constant is
included), Eq.$\;$(\ref{teleeq2}) does not compel the invariant
$\mathbb{S}\cdot \mathbb{T}$ to be null nor constant for vacuum
solutions, which raises the hope that a deformed teleparallelism
could be useful to smooth singularities of vacuum general
relativity solutions.

\section{Born-Infeld gravity}

Born-Infeld (\textbf{BI}) electrodynamics \cite{Born} has
experienced a renewed interest in the last years due to its close
connection with string theory, particularly because its capability
to describe the electromagnetic fields of D-branes \cite{Tsey},
\cite{Gaunt}. Inspired in these fruitful properties, together with
the ability of \textbf{BI} program concerning the cure of
singularities, we shall study a teleparallel theory of gravity
deformed \`{a} la Born-Infeld. In a rather different approach,
this subject has received some attention in the past
\cite{deser3,feingenbaum1,feingenbaum2,comelli1,comelli2,nieto},
where several deformations \`{a} la Born-Infeld combining higher
order invariants constructed with the curvature in a Riemannian
context were tried. All these constructions, however, lead to
troublesome four order field equations for the metric. As a matter
of fact, explicit solutions in four dimensions within these
frameworks were never found \footnote{In Ref. \cite{comelli2}, the
authors have considered exact solutions in two dimensions under
the assumption of maximally symmetric spacetimes, while in
\cite{feingenbaum1} numerical black holes solutions in four
dimensions were explored.}. In a different direction, \textbf{BI}
actions were explored more recently in Ref. \cite{Vollick,Banados}
using the Palatini formalism, where metric and connection are
taken as independent entities. In turn, following the lines of
\cite{Nos}, we will work with the Lagrangian
\begin{equation}
\mathcal{L}_{\mathbf{BI}}[e_{\mu }^{a}(x)]=\frac{\lambda \
e}{16\pi G}\left[ \sqrt{1+\frac{2\ (\mathbb{S}\cdot
\mathbb{T}-2\Lambda )}{\lambda }}-1\right], \label{bilagrangian}
\end{equation}%
where $\lambda $ is a constant that controls the scale at which
the deformed solutions depart from the original ones: Lagrangian
(\ref{bilagrangian}) tends to (\ref{lagrangianotel}) when
$\lambda\rightarrow\infty$. According to
Eq.$\;$(\ref{lagrangeeq}), the Euler-Lagrange equations become
\begin{eqnarray}
&&\partial _{\sigma }\left[ \left( 1+2\,\lambda
^{-1}(\mathbb{S}\cdot \mathbb{T}-2\Lambda )\right) ^{-1/2}\ e\
\,e_{a}^{\lambda }\ S_{\lambda }^{\ \ \nu \sigma }\right]   \notag
\\ &&-\left( 1+2\,\lambda ^{-1}(\mathbb{S}\cdot
\mathbb{T}-2\Lambda )\right) ^{-1/2}e\;e_{a}^{\lambda }\ \ S_{\eta
}^{\ \ \mu \nu }\,T_{\ \ \mu \lambda }^{\eta }  \notag \\
&&+\frac{\lambda }{4}\ e\;e_{a}^{\nu }\ \left[ \left( 1+2\,\lambda ^{-1}(%
\mathbb{S}\cdot \mathbb{T}-2\Lambda )\right) ^{1/2}-1\right]
\notag \\ &=&4\pi G\ e\,e_{a}^{\lambda }\ T_{\lambda }^{\ \nu }.
\label{biteleq}
\end{eqnarray}%
In order to explore the aptitude of deformed teleparallelism to
modify solutions of general relativity, we will try two types of
examples: the BTZ black hole and a n-dimensional cosmological
model with matter. In the first example, both \textbf{GR} and
teleparallel Lagrangians result to be constant for the chosen
solution, thus the deformation is limited to a shift of the
cosmological constant. In spite of this, teleparallelism exhibits
a better aptitude to deform the solution because it allows for a
BTZ solution even for positive cosmological constant. The strength
of modified teleparallelism is however revealed in solutions with
sources, where modified teleparallelism is able to control the
growing of the Hubble parameter by avoiding that the universe
reaches a singularity in a finite time.

\subsection{Extended BTZ black hole}

BTZ black hole is a vacuum solution for general relativity with
negative cosmological constant $\Lambda $ in 2+1 dimensions
\cite{btz}. The spinning BTZ metric is
\begin{eqnarray}
&&ds^{2}=\Big(-M-\Lambda \,r^{2}+\frac{J^{2}}{4r^{2}}\Big)\ dt^{2}
\label{btzinterval} \\
&&-\Big( -M-\Lambda \,r^{2}+\frac{J^{2}}{4r^{2}}\Big)^{-1}dr^{2}-r^{2}\Big(-%
\frac{J}{ 2r^2}\ dt+d\phi \Big)^{2},  \notag
\end{eqnarray}
where $M$ and $J$ are integration constants related to the mass
and the angular momentum respectively. For $\Lambda =-\ell ^{-2}$,
$M>0$ and $\mid J\mid \leq M\ell $ this metric has the structure
of a rotating black hole. The BTZ black hole displays event
horizons (the place where the lapse function vanishes) at
\cite{bthz}
\begin{equation}
r^{\pm }=\ell \ \left[ \frac{M}{2}\pm \frac{M}{2}\sqrt{\ 1-\left( \frac{J}{%
M\ell }\right) ^{2}}\right] ^{1/2},  \label{horizon}
\end{equation}%
and the ergosphere (the place where $g_{tt}$ vanishes) at
\begin{equation}
r^{erg}=\ell \ M^{1/2}>r^{+}>r^{-}.  \label{ergosphere}
\end{equation}%
The extremal case $\mid J\mid =M\ell $ corresponds to
$r^{+}=r^{-}=r_{erg}/\sqrt{2}$. A suitable dreibein field for the
metric (\ref{btzinterval}) is given by
\begin{eqnarray}
e^{0} &=&\Big(-M-\Lambda \,r^{2}+\frac{J^{2}}{4r^{2}}\Big)^{1/2}dt  \notag \\
e^{1} &=&\Big(-M-\Lambda \,r^{2}+\frac{J^{2}}{4r^{2}}\Big)^{-1/2}dr
\label{btzvielbein} \\
e^{2} &=&-\frac{J}{2r}\,dt+r\,d\phi.  \notag
\end{eqnarray}%
This dreibein satisfies Eq.$\;$(\ref{teleeq1}) for vanishing
energy-momentum tensor, and $\eta _{ab}\,e^{a}\,e^{b}$ reproduces
the interval (\ref{btzinterval}). Let us investigate how this
solution is affected by a deformation of the theory. In order to
understand the changes that the dreibein (\ref{btzvielbein}) has
to undergo for becoming a solution of the deformed equations
(\ref{biteleq}), let us note that the invariant $\mathbb{S}\cdot
\mathbb{T}$ results to be constant for the dreibein
(\ref{btzvielbein}): its value is $-2\Lambda $. Although
$\mathcal{L}_{\mathbf{T}} $ is not zero, a vacuum solution like
(\ref{btzvielbein}) which renders $\mathcal{L}_{\mathbf{T}}=$
\textit{constant} is very close to a vacuum solution of the
deformed theory. In fact, let us modify solution
(\ref{btzvielbein}) by replacing $\Lambda $ with a new constant
$\widetilde{\Lambda }$. Then $\mathbb{S}\cdot \mathbb{T}=-2\
\widetilde{\Lambda }$, so Eq.$\;$(\ref{biteleq}) turns out to be
\begin{eqnarray}
\partial _{\sigma }\left( e\ \,e_{a}^{\lambda }\ S_{\lambda }^{\ \ \nu
\sigma }\right) -e\;e_{a}^{\lambda }\ \ S_{\eta }^{\ \ \mu \nu
}\,T_{\ \ \mu \lambda }^{\eta }  \notag \\ +\frac{1}{4}\
e\;e_{a}^{\nu }\ \ \Bigg[ \mathbb{S} \cdot \mathbb{T}-2\ (2\
\Lambda +\widetilde{\Lambda })+\lambda  \notag \\ -\lambda \left(
1-4\,\lambda ^{-1}\,(\Lambda +\widetilde{\Lambda })\right) ^{1/2}
\Bigg] =0.  \label{biteleq2}
\end{eqnarray}
Since the solution we are trying solves the teleparallel vacuum equation (%
\ref{teleeq1}) for $\Lambda =\widetilde{\Lambda }$, then it will solve Eq.$\;$(%
\ref{biteleq2}) if $\widetilde{\Lambda }$ is chosen such that
\begin{equation}
-2\ (2\ \Lambda +\widetilde{\Lambda })+\lambda -\lambda \left(
1-4\,\lambda ^{-1}\,(\Lambda +\widetilde{\Lambda })\right)
^{1/2}=-2\ \widetilde{\Lambda }, \label{eqfora}
\end{equation}
i.e.,
\begin{equation}
\widetilde{\Lambda }=\Lambda \ (1-\epsilon )\
,\hspace{1cm}\epsilon =4\Lambda /\lambda.  \label{solutionfora}
\end{equation}
This solution represents a black hole if the effective
cosmological constant $\widetilde{\Lambda }$ is negative.
Summarizing, the BTZ dreibein for the deformed gravity theory
described by Lagrangian (\ref{bilagrangian}) is
\begin{eqnarray}
e^{0} &=&\Big(-M-\Lambda (1-\epsilon )\,r^{2}+\frac{J^{2}}{4r^{2}}\Big)%
^{1/2}dt  \notag \\
e^{1} &=&\Big(-M-\Lambda (1-\epsilon )\,r^{2}+\frac{J^{2}}{4r^{2}}\Big)%
^{-1/2}dr  \label{bibtzvielbein} \\
e^{2} &=&-\frac{J}{2r}\,dt+r\,d\phi ,  \notag
\end{eqnarray}
and the metric is
\begin{eqnarray}
ds^{2}=\Big(-M-\Lambda \ (1-\epsilon
)\,r^{2}+\frac{J^{2}}{4r^{2}}\Big)\ dt^{2}  \notag \\
-\Big(-M-\Lambda \ (1-\epsilon )\,r^{2}+\frac{J^{2}}{4r^{2}}\Big)
^{-1}dr^{2} \notag \\ -r^{2}\Big(-\frac{J}{2r^2}\ dt+d\phi
\Big)^{2}.  \label{bibtzinterval}
\end{eqnarray}
The dreibein (\ref{bibtzvielbein}) or the metric
(\ref{bibtzinterval}) genuinely differs from (\ref{btzvielbein})
and (\ref{btzinterval}); these two
metrics are not connected by a coordinate change because the invariant $%
\mathbb{S}\cdot \mathbb{T}$ is different for each one (also $R$ is
different). For negative effective cosmological constant
$\widetilde{\Lambda }=-\widetilde{\ell }^{-2}$ the solution is a
rotating BTZ black hole. Thus, even for $\Lambda >0$ the
\textbf{BI} Lagrangian (\ref{bilagrangian}) allows for BTZ
rotating black holes; specifically, the metric
(\ref{bibtzinterval}) is a
rotating black hole for $\Lambda <0$ and $\epsilon <1$, but also for $%
\Lambda >0$ and $\epsilon >1$. The horizons are placed at
\begin{eqnarray}
&&r_{BI}^{\pm }=\frac{r_{BI}^{erg}}{2}\left[ 1\pm \sqrt{\
1+\Lambda \ (1-\epsilon )\left( \frac{J}{M}\right) ^{2}}\right]
^{1/2}, \notag \\ &&r_{BI}^{erg}=\sqrt{\frac{M}{-\Lambda \
(1-\epsilon )}}. \label{bihorizon}
\end{eqnarray}
Let us compare the lengths of the horizons for the solutions (\ref%
{btzinterval}) and (\ref{bibtzvielbein}) corresponding to fixed
values of $\Lambda$, $M$ and $J$. Since the horizons are circles,
we will study the ratio $r_{BI}^{\pm }/r^{\pm }$. For
$\widetilde{\Lambda}<0$, three ranges of the parameter $\epsilon$
can be distinguished in this comparison:

\begin{itemize}
\item {\ \emph{Type I: $\epsilon <0\;($}}$\Lambda <0,\lambda >0)$. It
results $r_{BI}^{erg}/r^{erg}<1$, $r_{BI}^{+}/r^{+}<1$ and $%
r_{BI}^{-}/r^{-}>1$; then the horizons approach each other as a
consequence of the deformation.

\item {\emph{Type II: $\epsilon >1\;($}}$\Lambda >0)$. There is no black
hole for the \textbf{GR} counterpart.

\item {\emph{Type III: $0<\epsilon <1\;($}}$\Lambda <0,\lambda <0)$. It
results $r_{BI}^{erg}/r^{erg}>1$, $r_{BI}^{+}/r^{+}>1$ and $%
r_{BI}^{-}/r^{-}<1$; in this case the horizons move away from each
other as a consequence of the deformation. However the case
$\lambda <0$ will be rejected in next section since it produces
physically unacceptable solutions in cosmology.
\end{itemize}

\bigskip

This example shows the strategy to be followed to obtain deformed
solutions when one starts from a Lagrangian having a
``cosmological constant" term like the one in
Eq.$\,$(\ref{lagrangianotel}), i.e. $\mathcal{L}\propto
e\,(L-2\Lambda)$. If a given (vacuum) solution makes $L$ a
($\Lambda $ depending) constant, then replace $\Lambda $ in the
solution by a new constant $\widetilde{\Lambda }$ and substitute
the so built solution
in the modified field equation. Using that $L$ is constant, Eq.$\;$(\ref%
{lagrangeeq}) becomes
\begin{eqnarray}
\ ...-\partial _{\mu }\,\partial _{\nu }\left( e\ \frac{\partial \,L}{\ \
\partial \phi _{,\mu \nu }^{a}}\right) +\partial _{\mu }\left( e\ \frac{
\partial \,L}{\partial \phi _{,\mu }^{a}}\right)  \notag \\
-\ e\ \frac{\partial \,L}{ \partial \phi ^{a}}+\left(L-\frac{f(L-2\Lambda )}{%
f^{\ \prime }(L-2\Lambda ) } \right)\ \frac{\partial e}{\partial
\phi ^{a}}=0.
\end{eqnarray}
The proposed solution now solves the undeformed Euler-Lagrange equations for
cosmological constant $\widetilde{\Lambda}$. Therefore $\widetilde{\Lambda}$
should be chosen in such a way that
\begin{equation}
L(\widetilde{\Lambda})-\frac{f(L(\widetilde{\Lambda})-2\Lambda
)}{f^{\prime }(L(\widetilde{\Lambda})-2\Lambda )}= 2\
\widetilde{\Lambda},  \label{eqfora1}
\end{equation}
where $L(\widetilde{\Lambda})$ is the Lagrangian evaluated on the proposed
solution. This means that the deformation replaces the role of the
cosmological constant in the solution for a new parameter depending also on
the scale $\lambda $. Teleparallelism \`{a} la Born-Infeld (Lagrangian (\ref%
{bilagrangian})) uses the function $f$
\begin{equation}
f_{\mathbf{BI}}(x)=\lambda \sqrt{1+\frac{2\ x}{\lambda }}-\lambda,
\label{bifunction}
\end{equation}
so, writing Eq.$\;$(\ref{eqfora1}) for the function
(\ref{bifunction}), one gets Eq.$\;$(\ref{eqfora}) and the
solution (\ref{solutionfora}).

\bigskip

Einstein equations with cosmological constant imply
$L=-R=2\Lambda\,n/(n-2)$ for any vacuum solution in $n$ spacetime
dimensions. Therefore, vacuum solutions for theories
$f(-R-2\Lambda )$ can be straightforwardly obtained from general
relativity vacuum solutions by shifting $\Lambda$ to be
\begin{equation}
\frac{2\ n}{n-2}\ \widetilde{\Lambda} \ -\ \frac{f(2\widetilde{\Lambda}%
[n/(n-2)]-2\Lambda)}{f^{\prime}
(2\widetilde{\Lambda}[n/(n-2)]-2\Lambda)}\ =\ 2\
\widetilde{\Lambda}.  \label{rgsolutionfora}
\end{equation}
Contrasting with teleparallel Eq.$\,$(\ref{teleeq2}), the vacuum
solutions of \textbf{GR} share the same value of $L(\Lambda)$.
Thus, the effective cosmological constant (\ref{rgsolutionfora})
for modified \textbf{GR} is the same for all vacuum solutions.

Just for comparing with the Born-Infeld modified teleparallelism result (\ref%
{solutionfora}), let us compute the modified \textbf{GR} solution (\ref%
{rgsolutionfora}) for the Born-Infeld deformation (\ref{bifunction}) in $n=3$
dimensions. The result is
\begin{equation}
\widetilde{\Lambda }=\frac{\Lambda}{2}\left[ 1\
-\frac{1}{4\,\epsilon}\ \left(
1-\sqrt{1+8\,\epsilon}\right)\right].  \label{grconst}
\end{equation}
Thus the Born-Infeld deformation for \textbf{GR} is well defined
if $\epsilon>-1/8 $, and the bracket in Eq.$\;$(\ref{grconst}) is
positive definite. This means that the effective cosmological
constant keeps the sign of $\Lambda$. Therefore, in \textbf{GR}
modified \`{a} la Born-Infeld the BTZ black hole only exists for
$\Lambda<0$, which is a result different from the one obtained for
modified teleparallelism.

\subsection{Regular Cosmology}

The inability of Einstein-Hilbert Lagrangian to allow for high
energy deformed solutions not only embraces the vacuum solutions,
but any \textbf{GR} ($\Lambda =0$) solution satisfying $R=0$. This
assertion remains valid even if there are sources. For instance,
the Friedmann-Robertson-Walker (\textbf{FRW}) solution for a
radiation fluid cannot be smoothly deformed, because the
energy-momentum tensor is
traceless and so it is $R=0$. In contrast, the teleparellel Lagrangian (\ref%
{lagrangianotel}) does not vanish in this case; thus teleparallelism allows
for a smooth deformation of such kind of solution \cite{Nos}.

Let us study the deformation (\ref{bilagrangian}) in the context
of a spatially flat \textbf{FRW} geometry in the presence of a
homogeneous and isotropic fluid. Then the source is represented by
the stress-energy tensor $T_{\ \ \nu }^{\mu }=\emph{diag}(\rho
(t),-p(t),-p(t),...)$ in the comoving reference frame. The
teleparallel equations can be solved by considering the vielbein
\begin{equation}
e_{\mu }^{a}=\emph{diag}(1,\, a(t),\, a(t),\, ...),\ \ \ \ \ \
e=a^{n-1}, \label{frame}
\end{equation}
leading to the metric $g_{\mu\nu}=\emph{diag}(1,-a(t)^2,-a(t)^2,...)$. In
this case the only non null components of $\mathbb{S}$ and $\mathbb{T}$ are
\begin{eqnarray}  \label{sandt}
&&S_{\alpha 0\alpha }=-S_{\alpha \alpha 0}=-\frac{1}{2}(n-2)\ a(t)\ \dot{a}%
(t),  \notag \\
\\
&&T_{\alpha 0\alpha }=-T_{\alpha \alpha 0}=a(t)\ \dot{a}(t), \ \ \
\ \ \ \ \ \alpha \neq 0.  \notag  \label{componentes}
\end{eqnarray}
Thus $\mathbb{S}\cdot \mathbb{T}=-(n-1)(n-2)\ H^{2}$,
$H=\dot{a}(t)/a(t)$ being the Hubble parameter, which is not null
nor constant whenever a source
is present. The first term in the equation (\ref{biteleq}) for the indexes $%
a=0=\nu$ is null; then the \textit{initial value} equation for the
modified \textbf{FRW} cosmology results
\begin{equation}
\frac{1-\epsilon}{\left(
1-\epsilon-2(n-1)(n-2)\,\frac{H^{2}}{\lambda}\right)
^{\frac{1}{2}}}-1=\frac{16\pi G}{\lambda}\ \rho.  \label{valin}
\end{equation}
The isotropy of the proposed solution makes equal the equations (\ref%
{biteleq}) for spatial indexes $a=\nu$; they are
\begin{eqnarray}
&&(1-\epsilon)\left(2(n-2)\,q\,\frac{H^{2}}{\lambda} + 2n(n-2)\,\frac{H^{2}}{%
\lambda}-1+\epsilon\right)  \label{esp} \\
&&\times\left( 1-\epsilon-2(n-1)(n-2)\,\frac{H^{2}}{\lambda}\right) ^{-\frac{%
3}{2}}+1\ =\ \frac{16\pi G}{\lambda}\ p\ \ \ \ \   \notag
\end{eqnarray}
In the last expression $q=-a\dot{a}^{-2}\ddot{a}=-(1+\dot{H}\
H^{-2})$ is the deceleration parameter. Eqs.$\;$(\ref{valin}),
(\ref{esp}) lead to the energy-momentum conservation. In fact, by
differentiating the initial value equation (\ref{valin}) with
respect to the time, and combining this result with
Eq.$\;$(\ref{esp}), one gets
\begin{eqnarray}
\frac{d}{dt}\left( \rho \ a^{n-1}\right) =-p\ \frac{d}{dt}a^{n-1},\hspace{%
0.5in}\ \ \ \mathrm{or}  \notag \\
\dot{\rho}+(n-1)\ (\rho +p)\ H=0.  \label{densidad}
\end{eqnarray}
For a barotropic fluid satisfying the state equation $p=\omega(n)
\rho$, the conservation law (\ref{densidad}) leads to the behavior
\begin{equation}
\rho (t)=\rho _{o}\ \left( \frac{a_{o}}{a(t)}\right)
^{(n-1)(1+\omega )}. \label{rodet}
\end{equation}

Equations (\ref{valin})-(\ref{esp}) in vacuum ($\rho =p=0$) have the
solution $H=\pm H_{o}$, $q=-1$, for the constant $H_{o}^{\;2}= 2\Lambda
(1-\epsilon )/[(n-1)(n-2)]$. In this case the result is the de Sitter metric
for the effective cosmological constant $\widetilde{\Lambda }=\Lambda \
(1-\epsilon )$. The similarity with the shift of the previous section comes
from the fact that the invariant is $\mathbb{S} \cdot \mathbb{T}=-2\
\widetilde{\Lambda }$ in both cases.

In the presence of a barotropic fluid, the system (\ref{valin}), (\ref{rodet}%
) can be rewritten by using the variable
\begin{equation}  \label{variabley}
y=\ln\left[(\frac{a}{a_o})^{(n-1)(1+\omega)}\right]\ \Rightarrow\ \dot{y}%
=(n-1)(1+\omega)\,H.
\end{equation}
Thus the dynamics of the spatially flat \textbf{FRW} universe in
Born-Infeld teleparallelism is described by the equation
\begin{eqnarray}
\frac{2(n-2)}{(n-1)(1+\omega)^2}\ \dot{y}^2&+&\frac{\lambda\ (1-\epsilon)^2}{%
(1+16 \pi G \rho _{o}\lambda^{-1} \exp(-y))^2}  \notag \\
&& =(1-\epsilon)\ \lambda,  \label{bardo}
\end{eqnarray}
whose \textbf{GR} ($\lambda\rightarrow\infty$) limit is
\begin{equation}
\frac{2(n-2)}{(n-1)(1+\omega)^2}\ \dot{y}^2-32\pi G\rho_o\exp(-y)
=4\,\Lambda. \label{bardo2}
\end{equation}

\bigskip

The variable $y$ is monotone increasing with the scale factor
$a(t)$. So the behavior of the scale factor can be read directly
in the ``energy conservation" equations (\ref{bardo}) and
(\ref{bardo2}). As known, the effective potential for a spatially
flat \textbf{FRW} universe in \textbf{GR} expands forever for
$\Lambda>0$ and recollapses for $\Lambda<0$. On the other hand,
the Born-Infeld teleparallel potential for $\lambda>0$ is an
increasing function, vanishing for $y\rightarrow -\infty$
($a\rightarrow 0$) and going to $\lambda(1-\epsilon)^2$ for
$y\rightarrow\infty$. Since the energy level in
Eq.$\;$(\ref{bardo}) is $\lambda(1-\epsilon)$, then: {I)} the
universe recollapses if $1-\epsilon>1$ (i.e. $\Lambda<0$), or
{II)} expands forever if $0<1-\epsilon<1$ (i.e.
$0<\Lambda<\lambda/4$). Although this behavior seems not to differ
considerably from the \textbf{GR} one, it should be emphasized
that the main difference lies in the behavior of the Hubble
parameter when $y\rightarrow -\infty$: while $H$ diverges in
\textbf{GR}, in Born-Infeld teleparallelism $H$ goes to the
constant value
\begin{equation}  \label{Hmax}
H^2\rightarrow \frac{(1-\epsilon)\ \lambda}{2(n-1)(n-2)}=\frac{%
\lambda-4\Lambda}{2(n-1)(n-2)}.
\end{equation}
For $\lambda<0$ and $\epsilon\neq1$ the effective potential
becomes an infinite well. This is an unphysical feature, since it
leads $H$ to diverge for $16\pi G\rho=|\lambda|$ (see Eq.$\,$
(\ref{valin})). Therefore we will only consider the case
$\lambda>0$.

The dependence on time of the scale factor can be obtained from
the initial value equation (\ref{valin}) or, equivalently, from
(\ref{bardo}). We will use the variable
\begin{equation}\label{variabley1}
\mathtt{y}=\frac{\lambda}{16\, \pi\, G\, \rho_o}\
\left(\frac{a(t)}{a_o}\right)^{(n-1)(1+\omega)}.
\end{equation}
In this way, the initial value equation takes the form

\begin{equation}  \label{integral}
\dot{\mathtt{y}}= \pm \mathcal{A}\
\frac{\mathtt{y}}{1+\mathtt{y}}\, \sqrt{1+2\,
\mathtt{y}+\epsilon\,\mathtt{y}^2},
\end{equation}
with $\mathcal{A}=(1+\omega)\sqrt{\frac{\lambda(1-\epsilon)(n-1)}{2(n-2)}}$
a non null constant. The solution of (\ref{integral}) can be obtained in a
closed but implicit form by direct integration, and depends on the sign and
range of the parameter $\epsilon$. Concretely, we have two types of
solutions:

\begin{itemize}
\item {\emph{Type I: $\epsilon<0$ $(\Lambda <0)$}} (the universe
recollapses)
\begin{equation}  \label{caso1}
\pm \mathcal{A}\,t\pm c=\mathcal{F}(\mathtt{y})-\frac{1}{%
(-\epsilon)^{1/2}} \arcsin\Big[\frac{1+\epsilon\,\mathtt{y}}{\sqrt{%
1-\epsilon}}\Big].
\end{equation}

\item {\emph{Type II: $0<\epsilon<1$ $(\Lambda >0)$}} (the
universe expands forever)
\begin{equation}  \label{caso2}
\mathcal{A}\,t+c = \mathcal{F}(\mathtt{y}) +\frac{1}{\sqrt{\epsilon}}\ln\Big[%
\frac{1+\epsilon\,\mathtt{y}}{\sqrt{\epsilon}}+\sqrt{1+2\,
\mathtt{y}+\epsilon\,\mathtt{y}^2}\Big],
\end{equation}

\end{itemize}
where the function $\mathcal{F}(\mathtt{y})$ is in both cases
\begin{equation}
\mathcal{F}(\mathtt{y})=\ln\Big[\frac{\mathtt{y}}{1+\mathtt{y}+\sqrt{1+2\,
\mathtt{y}+\epsilon\,\mathtt{y}^2}}\Big].
\end{equation}

In Eq.$\,$(\ref{caso1}) the sign $\pm$ corresponds to the
expanding and the collapsing branch respectively. Both branches
can be joined at $t=0$ by choosing the integration constant $c$ to
equalize the right member for the maximum scale factor. According
to Eq.$\,$(\ref{integral}) the maximum scale factor
($\dot{\mathtt{y}}=0$) is
\begin{equation}\label{ymax}
\mathtt{y}_{max}=\frac{1+\sqrt{1-\epsilon}}{-\epsilon},
\end{equation}
thus
\begin{equation}
c=
-\ln\Big[1-\frac{\epsilon}{1+\sqrt{1-\epsilon}}\Big]+\frac{\pi/2}{(-\epsilon)^{1/2}}.
\end{equation}

\begin{figure}[ht]
\centering
\includegraphics[scale=.75]{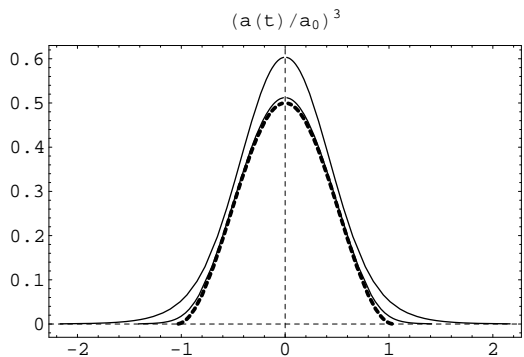} \caption[]{Behavior of the non-dimensional
cubed scale factor $(a(t)/a_{0})^3$ as emerges from (\ref{caso1})
for $\omega=1/2$, $\Lambda=-1$ in $n=3$ dimensions. The upper
curve is for $\epsilon=-1$, the middle curve is for
$\epsilon=-0.1$ and the dashed one corresponds to
\textbf{GR}.}\label{Fac}
\end{figure}

Figure \ref{Fac} shows the \emph{Type I} recollapsing case. The
scale factor as a function of time is depicted for a radiation
fluid in three dimensions, i.e. $\omega=1/2$ and $n=3$. Besides we
set $16\, \pi\, G\, \rho_o=1$ and $\Lambda=-1$. The upper, middle
and lower (dashed) curves correspond to $\epsilon=-1$
($\lambda=4$), $\epsilon=-0.1$ ($\lambda=40$), and \textbf{GR}
($\lambda\rightarrow\infty$) respectively. Note that the
\textbf{GR} scale factor exists only for $-1\leq t \leq1$, whereas
it exists for all values of time in the \textbf{BI} case.

Physically more relevant, at least in four dimensions, is the
\emph{Type II} case where the cosmological constant is positive.
In this case, the Eq.$\;$(\ref{caso2}) says that the late time
behavior ($\mathtt{y}\rightarrow \infty$) of the scale factor is
$a(t)\propto\exp[\sqrt{\frac{2\Lambda(1-\epsilon)}{(n-1)(n-2)}} \,
t]$, while the initial stage is described by $a(t)\propto
\exp[\sqrt{\frac{\lambda(1-\epsilon)}{2(n-1)(n-2)}}\, t]$ (see
Eq.$\;$(\ref{Hmax})). Thus, the universe evolves from an
inflationary stage, driven by the (vacuum-like) energy
$\lambda(1-\epsilon)$, to another exponential epoch ruled by the
vacuum energy $\Lambda(1-\epsilon)$ (a similar de Sitter-de Sitter
evolution was obtained in a quite different approach in Ref.
\cite{Saulo}). Since $\epsilon$ should be very small in order that
the theory does not appreciably differ from \textbf{GR} for most
of the history of the universe (see a lower bound for $\lambda$ in
Ref.$\;$\cite{Nos}), one concludes that in four dimensions the
scale factor evolves in time as
\begin{equation}\label{esquema}
a(t\rightarrow-\infty)\propto e^{\sqrt{\frac{\lambda}{12}}\,
t}\,\,\,\rightsquigarrow\,\,\,a(t\rightarrow\infty)\propto
e^{\sqrt{\frac{\Lambda}{3}} \, t}.
\end{equation}
Finally, the limiting case $\epsilon=1$ corresponds to the scale
factor being constant, as follows from Eq.$\,$(\ref{bardo}).

\section{Concluding Comments}
In spite of having different causal structures, the BTZ black hole
is locally the anti-de Sitter spacetime \cite{bthz}, i.e. the
maximally symmetric solution with negative cosmological constant.
When evaluated on maximally symmetric solutions, both $R$ and
$\mathbb{S}\cdot \mathbb{T}$ Lagrangians are equal to a constant
that is independent of the integration constants: it only depends
on $\Lambda$. So, in both theories, general relativity and
teleparellelism, the deformation of these maximally symmetric
solutions just amount to a shifting of the cosmological constant.
The shifting is controlled by the non-dimensional parameter
$\epsilon= 4\Lambda/\lambda$, where $\lambda$ is a
Born-Infeld-like constant going to infinity when the deformed
theory approaches the original one. The cosmological constant
$\Lambda$ and the (shifted) effective cosmological constant
$\widetilde{\Lambda }$ can have opposite sign in deformed
teleparallelism; so the anti-de Sitter solution can solve the
deformed teleparallel equations even for positive cosmological
constant.

On the other hand, we have studied the deformation of non-vacuum
cosmological solutions. Although the parameter $\epsilon$ takes
part in Eq.$\,$(\ref{integral}), its presence does not alter the
non-deformed result that spatially flat Friedmann-Robertson-Walker
non-vacuum solutions expand for $\Lambda>0$ and recollapse for
$\Lambda<0$. Instead, the deformed theory smoothes the initial
singularities, which is the effect pursued by Born-Infeld
deformations. In fact, the Hubble parameter goes to a constant
when the scale factor $a$ vanishes (see Eq.$\,$(\ref{Hmax})). This
value is also the maximum value to be attained by $H$ (see
Eq.$\,$(\ref{valin})).

The \textbf{BI} approach (\ref{bilagrangian}) generates regular
solutions. In the cosmological setting this is so, not only
because the scale factor is always different from zero, but
because the geometrical invariants (both, in Riemann and
Weitzenb\"{o}ck spacetimes) are bounded for all finite proper
times. In fact, each invariant in Weitzenb\"{o}ck spacetime that
is quadratic in the torsion tensor has to be proportional to $H^2$
in the setting (\ref{frame}) (see Eq.$\;$(\ref{sandt})). On the
other hand, the Riemannian invariants for the metric
$g_{\mu\nu}=diag(1,-a(t)^2,-a(t)^2,...)$ can be cast in the
polynomical form $\mathcal{P}=(H,\dot{H})$. For instance, in four
dimensions, the scalar curvature is $R=6(2H^2+\dot{H})$, the
squared Ricci scalar $R_{\mu\nu}^2=R^{\mu\nu}R_{\mu\nu}$ is
$R_{\mu\nu}^2=12(3H^4+3H^2 \dot{H}+\dot{H}^2)$, and the
Kretschmann invariant $K=
R^\alpha_{\,\,\beta\gamma\delta}R_{\alpha}^{\,\,\beta\gamma\delta}$
reads $K=12 (2H^4+2 H^2 \dot{H}+ \dot{H}^2)$. All these invariants
are well behaved due to the saturation value (\ref{Hmax}) that the
Hubble parameter reaches as $a(t)\rightarrow 0$. Regarding this
matter, the time derivative of Eq.$\;$(\ref{integral}) combined
with the definition given in Eq.$\;$(\ref{variabley1}), shows that
\begin{equation}
\dot{H}=-\alpha \frac{\mathtt{y}^2}{(1+\mathtt{y})^3},
\end{equation}
where $\alpha=\lambda(1+\omega)(1-\epsilon)^2/2(n-2)$ is a non
null constant. By setting $n=4$, $\omega=1/3$ and $\epsilon=0$,
one finds the following expressions for the invariants
\begin{eqnarray}  \label{invgeom}
R &=& \lambda \Big[\frac{1+3\mathtt{y}}{(1+\mathtt{y})^3}\Big],
\notag \\ R_{\mu\nu}^2 &=&
\frac{\lambda^2}{12}\Big[\frac{3+18\mathtt{y}+ 27 \mathtt{y}^2+4
\mathtt{y}^4}{(1+\mathtt{y})^6}\Big],  \notag \\ K
&=&\frac{\lambda^2}{6}\Big[\frac{1+6\mathtt{y}+9 \mathtt{y}^2+4
 \mathtt{y}^4 }{(1+\mathtt{y})^6}\Big].
\end{eqnarray}
This means that the BI parameter $\lambda$ not only bounds the
dynamics of $H(t)$ and characterizes the minimum density for
having inflation \cite{Nos}, but also establishes a maximum
attainable curvature.

One could wonder if the \textbf{BI} framework here considered can
be viewed as a particular case of a more general determinantal
Born-Infeld action for gravity. Indeed, following in a closer way
the \textbf{BI} spirit, one could try the n-dimensional
determinantal action
\begin{equation}  \label{acciondet}
\mathcal{I}_{BIG}= \lambda\int d^n
x\Big[\sqrt{\det(g_{\mu\nu}+\lambda^{-1}F_{\mu\nu})}-\sqrt{\det(g_{\mu\nu})}\Big],
\end{equation}
where $F_{\mu\nu}$ is quadratic in the Weitzenb\"{o}ck torsion:
$F_{\mu\nu}= A \,S_{\mu\lambda\rho}T_{\nu}^{\,\,\,\lambda\rho}+B\,
S_{\lambda\mu\rho}T^{\lambda\,\,\,\, \rho}_{\,\,\,\,\,\nu}$, $A$
and $B$ being non-dimensional constants. Such a combination
ensures the correct \textbf{GR} limit since both constituents of
$F_{\mu\nu}$ have trace $\mathbb{S}\cdot\mathbb{T}$. Besides, the
dynamical equations coming from (\ref{acciondet}) will be still of
second order in the vielbein derivatives. For the choice $2A+B=0$,
the action (\ref{acciondet}) reproduces the solutions considered
in the last section, though the equivalence with the scheme
(\ref{bilagrangian}) for other solutions is not clear yet
\cite{Nos2}. The complete characterization of the theory
(\ref{acciondet}) for the whole parameter space $(A,B)$ will be
matter of future developments.
\bigskip
\acknowledgements
We would like to thank G. Giribet for the
constant encouragement afforded during this work. This research
was supported by CONICET.

\end{document}